\documentclass{revtex4}
\pdfoutput=1
\usepackage{slashed}
 \usepackage{epstopdf}
\usepackage{graphicx}
\begin{document}

\title{\bf POSITIVITY VIOLATIONS IN QCD}

\author{John M. Cornwall\footnote{Email cornwall@physics.ucla.edu  }}

\affiliation{Department of Physics and Astronomy, UCLA\\
Los Angeles, CA 90095\\
cornwall@physics.ucla.edu}

\begin{abstract}

Both lattice simulations and theoretical studies show that   the spectral function of the   gluon propagator of QCD (in various gauges, as well as for the gauge-invariant Pinch Technique, or PT, propagator) is not non-negative everywhere, although it should be   if it has a physical interpretation as in QED. Theory says moreover that the non-positive spectral function of the Landau-gauge or of the PT gluon propagator is further constrained to obey a superconvergence relation (the integral of the spectral function vanishes).  We review the  theoretical and lattice evidence for violation of positivity as well as various interpretations of this violation, and consider methods for checking superconvergence on the lattice (so far undone).  The most common interpretation is that positivity violation implies confinement of gluons, so the gluon propagator does not describe   processes with physical gluons.  Another more direct and gauge-invariant  interpretation arises from the PT:  Asymptotic freedom alone demands non-positivity and superconvergence.  

\end{abstract}
\maketitle

\section{Introduction}

All students of quantum field theory are taught that the propagator of a scalar field has a K\"allen-Lehmann representation
\begin{equation}
\label{klrep}
\Delta (p^2) =\frac{1}{\pi}\int_0^{\infty}\,\mathrm{d}\sigma \,\frac{\rho(\sigma)}{p^2-\sigma+\mathrm{i}\epsilon}
\end{equation}
with a non-negative spectral function $\rho (\sigma )$.  Above a certain threshhold, $\rho$ is positive, which seems to be required by the interpretation of $\rho$ as a sum of absolute squares.  

Things may be different when spin enters, but for the important case of QED it appears that the photon propagator has a non-negative $\rho$.  This happens because the photon propagator is gauge-invariant and $\rho$ has contributions only from physical and positive cross-sections.   Yet this is a red herring, since positivity comes about only at the cost of a fatal flaw in QED:  The ultraviolet Landau pole, at which the running charge becomes infinite and its square becomes negative for higher momenta.  

Turn now to asymptotically-free (AF) non-Abelian gauge theories (NAGTs) such as QCD, which are consistent theories of spin-one gluons because they have no UV Landau pole.  (In perturbation theory they have Landau poles in the infrared that have to be cured by some form of mass generation; we take this for granted.)  But the properties of the spectral function are not obvious.  Unlike the photon propagator, the usual gluon self-energy as defined by Feynman diagrams depends on the gauge chosen, and can have negative contributions to the spectral function.  And finally, to the extent that the gluon is not capable of existing as a free physical particle, one might question the very existence of a gluon propagator, let alone its meaning.

Extracting  the spectral function from lattice data is much more than an academic exercise; it is an important step toward constructing the Minkowski-space propagator, with its information on the nature of the (now well-established) gluon mass and other physical features.  But that extraction is by no means straightforward, as we will see, and its physical interpretation is obscure to the extent that the propagator is simulated in a non-physical gauge.  Ideally, one would simulate the gauge-invariant Pinch Technique\cite{corn076,cornbinpap} (PT) propagator, which means to simulate it, either directly or indirectly, in the background field method (BFM) Feynman gauge\cite{cornbinpap,papbin}, convert\cite{corn138,corn141,pap09} it to the PT-RGI propagator that is both gauge invariant and renormalization-group invariant (RGI) by a simple multiplication by the coupling $g^2$, and extract the resulting PT-RGI spectral function.  This program is in its very earliest stages, and in this review we concentrate mostly on what needs to be done to extract important properties of the spectral function.

\section{\label{afnonp} Asymptotic freedom and non-positivity}

As we review in Sec.~\ref{latt}, there is no doubt that the spectral function of an AF propagator in a covariant gauge such as Landau gauge has a spectral function with non-positive contributions.

In one way it is not surprising that an AF  gluon propagator violates positivity in a covariant gauge, because as conventionally described (by Feynman diagrams in perturbation theory) such a gauge for the propagator leads to states with negative norms.  But perhaps one should be surprised, because the same is true for the photon propagator in QED, yet we have no reason to believe that the QED spectral function has any negative parts.  This suggests that somehow AF makes the difference.  

Several authors made this point, beginning long ago\cite{corn076,oehzimm} and continuing to more recent times\cite{strauss,alksme}. Some authors\cite{oehzimm,strauss,alksme} study  the Landau-gauge propagator; another\cite{corn138} considers the gauge- and renormalization-group-invariant (RGI) propagator $\widehat{\Delta}$ of the pinch technique\cite{corn076} (PT).  We call this the PT-RGI propagator\cite{corn141}, and discuss it briefly in Sec.~\ref{ptsec}.   From the PT-RGI point of view the lack of positivity has a natural and physical interpretation for an AF theory\cite{corn138}, essentially coming from the logarithmic vanishing of the running charge.   Although the Landau-gauge propagator is not in itself a physical object and has no such interpretation, Oehme and Zimmermann's results are important for comparison with the Landau-gauge propagator that is measured on the lattice.  On the other hand, any PT-RGI function should contain physical information, and the PT-RGI propagator certainly does.  But it is not like QED, where the positive spectral function can be, in principle, reconstructed from physical cross-sections.  

Even though there are presently no techniques for directly simulating the PT-RGI gluon propagator on the lattice, it may be possible\cite{pap09} to retrieve this from Landau-gauge simulations as discussed in Sec.~\ref{non-pos}.

\subsection{\label{non-pos} Non-positivity for the Landau-gauge propagator}

We now switch to a Euclidean representation, with metric $p^2=\sum p_i^2$.  
One can show\cite{oehzimm} from general principles such as the renormalization group and AF that the Landau-gauge gluon propagator has a non-positive spectral function.  Write this propagator as:
\begin{equation}
\label{landprop}
\Delta_{\mu\nu}(p^2)=P_{\mu\nu}(p)\Delta (p^2);\;P_{\mu\nu}(p)=\delta_{\mu\nu}-\frac{p_{\mu}p_{\nu}}{p^2}
\end{equation}
with 
\begin{equation}
\label{spect}
\Delta (p^2)=\frac{1}{\pi}\int_0^{\infty}\!\mathrm{d}\sigma \!\frac{\rho(\sigma)}{p^2+\sigma}
\end{equation}
and $p^2\geq 0$.  We do not explicitly call out the perturbative zero-mass pole, corresponding to a contribution $\rho (\sigma )\sim \delta (\sigma )$.  The renormalization group says that at asymptotically-large momentum the propagator has the behavior
\begin{equation}
\label{afozimm}
\Delta (p^2)\sim \frac{1}{p^2}[\ln (\frac{p^2}{\mu^2})]^{-\alpha_0/2b}
\end{equation}
where $\alpha_0g^2$ is the leading term of the anomalous dimension of the propagator in the Landau gauge, $-bg^3$ is the lowest-order term in the beta-function for coupling constant $g$, and $\mu$ is the renormalization-point mass.  One sees already that trouble lies ahead for a physical interpretation, since $\alpha_0$ depends on the gauge (of course, $b$ does not).  In the Landau gauge, and with no matter fields, one finds a positive value:
\begin{equation}
\label{ratio}
\frac{\alpha_0}{2b}=\frac{13}{22}.
\end{equation}
The Landau-gauge propagator therefore vanishes faster than $1/p^2$ near infinity, which manifestly contradicts the representation (\ref{spect}) if $\rho$ is positive.    Clearly, from (\ref{spect}) if $\Delta$ vanishes faster than $1/p^2$ at infinity, a superconvergence relation
\begin{equation}
\label{supercon}
\int\,\mathrm{d}m^2\,\rho (m^2)=0
\end{equation}
holds and there must be some region  where $\rho$ changes sign.    We take the usual convention that the tree-level propagator is $+1/p^2$ meaning that $\rho$ must be negative for asymptotically-large $\sigma$.    
Given the asymptotic behavior of the gluon propagator:
\begin{equation}
\label{largep}
\Delta (p^2)\sim \frac{1}{p^2(\ln p^2)^A}
\end{equation}
the spectral function behaves like:
\begin{equation}
\label{specfunc}
\rho (\sigma )\sim \frac{-\pi A}{\sigma(\ln \sigma )^{1+A}}
\end{equation}

This conclusion puzzles Oehme and Zimmermann, who expect a positive $\rho$ and therefor a {\em negative} value for $\alpha_0/(2b)$.  This is the situation of QED, but this negative value comes with a UV Landau pole where the running charge blows up.  These authors argue that for QCD perhaps the Landau gauge is not representative of the true physics, since the other possibility for changing the sign while retaining AF is that the number of quark flavors lies between 10 and 16.

Another possibility might be that an integral representation such as (\ref{spect}) does not hold, even with a non-positive spectral function, and that the propagator is an entire function. But the authors go on to show that the gluon propagator cannot be an entire function, except at an infrared-stable fixed point (which ordinary QCD does not have), and must have a cut extending to infinity.

\subsection{\label{ptprop}  Non-positivity for the PT-RGI propagator}

The PT leads to a qualitatively-similar conclusion\cite{corn076,corn138,corn141} but an entirely different interpretation, discussed in Sec.~\ref{ptsec}.  For the PT-RGI propagator the anomalous dimension $\alpha_0$ is precisely $2b$, independent of the gauge, so that the asymptotic behavior of the PT-RGI propagator $\widehat{\Delta}$ is
\begin{equation}
\label{ptasym}
\widehat{\Delta}(p)\rightarrow \frac{1}{b p^2 \ln (\frac{p^2}{\Lambda^2})} 
\end{equation}
where $\Lambda$ is the physical and gauge-invariant renormalization-group mass, not the renormalization-point mass.  We see here the appearance of the lowest-order running charge:
\begin{equation}
\label{runch}
\bar{g}^2(p^2)=\frac{1}{b\ln (\frac{p^2}{\Lambda^2})}
\end{equation}
so that 
\begin{equation}
\label{ptasym2}
\widehat{\Delta}(p)\rightarrow \frac{\bar{g}^2(p^2)}{p^2},
\end{equation}
a product of the running charge and a normal free propagator.   The PT gluon propagator vanishes faster than $1/p^2$, just as the Landau-gauge propagator does, and it has a spectral representation of the form of Eq.~(\ref{spect}) with the spectral function satisfying the superconvergence relation  (\ref{supercon}) with a non-positive spectral function, corresponding to $A=1$ in the asymptotic behavior of Eq.~(\ref{specfunc}).  

 To all orders and all momenta\cite{papbin,cornbinpap} the Landau-gauge propagator and the PT-RGI propagator differ by a factor:
\begin{equation}
\label{cbpeq}
\widehat{\Delta}(p)= \frac{g^2\Delta (p^2)}{[1+G(p^2)]^2}.
\end{equation}
where $G(p^2)$ is a complicated  Green's function, depending on the gauge, on the coupling $g$, and on the renormalization-point mass.    Needless to say, the asymptotic behavior of $G$ is just what is needed to reconcile Eqns.~(\ref{afozimm},\ref{ptasym}) in lowest order.  Eq.~(\ref{cbpeq}) is an example of a background-quantum identity, relating conventional Green's functions to BFM Green's functions in a given gauge, such as Landau gauge\cite{papbin,cornbinpap,ghs}.  
There are   direct lattice simulations\cite{stern2} and PT-based theoretical studies\cite{pap09} of $G(q^2)$.  The latter yields the first approximate estimate tied to lattice data of the PT-RGI propagator.   These simulations, depending on the so-called ghost dressing function, show that the Kugo-Ojima\cite{ko} confinement criterion   $G(0)=-1$ is violated.

\section{\label{latt} Lattice results on non-positivity and superconvergence}

Lattice simulations are always done for Euclidean (spacelike, in Minkowski terms) momenta, which means that some vital information about the gluon propagator is not available directly, but must be inferred from necessarily approximate considerations.  This propagator is real in Euclidean space, so it is not a simple thing to extract the spectral function.  Of course, in a Minkowski-space integral such as Eqn.~(\ref{klrep}) one need only take the imaginary part, but in Euclidean space   the spectral function itself emerges only after an inverse Laplace transformation requiring continuation into the entire complex plane ( see Eq.~(\ref{wallcor}) below), and this is an ill-posed problem, difficult to solve accurately, as we will document in Sec.~\ref{lattdata}.  Unfortunately, it is rather easy to guess at quite different spectral functions that give (within the simulation accuracy) the same Euclidean propagator.  Fortunately, one can answer the essentially qualitative question of whether a spectral function is positive or not by looking at prominent global properties of these integral transformations.  To be quantitative is a different matter.  In particular, to verify superconvergence on the lattice is difficult, and not yet attempted directly.  Some authors have argued for superconvergence from lattice data, both for the Landau gauge\cite{iritani} and for the maximal Abelian gauge\cite{gongyo}.  In both cases the authors fit the gluon propagator to a simple function that is trivially extensible into the complex plane and that has superconvergence built in.  We discuss here alternatives based on the standard lattice tool of the Fourier transform in time of the propagator at zero spatial momentum.   

\subsection{\label{analysis}  Analytic preliminaries}

The Fourier transform of interest is often called the wall-to-wall correlator $C(t)$;  Mandula has given\cite{mand} a brief but useful review.  This correlator clearly reveals non-positivity through its qualitative behavior at fairly large values (in lattice units) of $t$, where it should differ only slightly from its continuum value.  To find superconvergence requires a much more quantitative study for small $t$, as we show below.

With the notation of Eqs.~(\ref{landprop},\ref{spect}), $C(t)$ is defined by
\begin{equation}
\label{wallcor}
C(t)=\frac{1}{2}\int_{-\infty}^{\infty}\,\mathrm{d}p\,e^{ipt}\Delta (p^2)=\int_0^{\infty}\,\mathrm{d}m\,\rho(m^2)e^{-mt}.
\end{equation}
Define also the {\em continuum} effective mass function
\begin{equation}
\label{effmass}
M_{eff}(t)=-\frac{d}{dt}\ln C(t)\equiv \langle m\rangle
\end{equation}
which defines the measure.
For a free particle of non-running mass $m_0$ the effective mass is precisely $m_0$.  But for a general spectral function $M_{eff}$ is a non-trivial function of $t$.  
On a lattice of lattice spacing $a$ it is canonical to define the effective mass function as
\begin{equation}
\label{latteffmass}
M_{eff-l}=-\lim_{a\to 0}\frac{1}{a}\ln [\frac{C(t+a)}{C(t)}] = -\frac{\dot{C}(t)}{C(t)}. 
\end{equation}
We can also define the equivalent of $\dot{M}_{eff}(t)$ on the lattice through
\begin{equation}
\label{mlattdot}
-\dot{M}_{eff-l}(t)=(\frac{d}{dt})^2\ln C(t)=\lim_{a\to 0} \frac{1}{a^2}\ln [\frac{C(t+2a)C(t)}{C^2(t+a)}]=[\frac{\ddot{C}}{C}
-(\frac{\dot{C}}{C})^2].
\end{equation}
Of course in practice the limits are unattainable, and one has to extrapolate data at finite lattice spacing.

\subsubsection{Discovering non-positivity on the lattice}

If $\rho$ is non-negative, it easily follows from Eq.~(\ref{spect}) that the derivatives of $\Delta (p^2)$ alternate in sign  and  the same is true for the derivatives of $C(t)$,    which are proportional to the moments $\langle m^N\rangle$.  The absolute values of the derivatives of $\Delta$ monotonically decrease.  For   a non-negative $\rho$ we see that $C(t)$ is analogous to a statistical-mechanics partition function at temperature $1/t$.  The connected moments of $C(t)$ come from derivatives of $\ln C$; for example,  
\begin{equation}
\label{sineq}
(\frac{d}{dt})^2\ln C(t)= -\frac{dM_{eff}(t)}{dt} = \langle m^2\rangle -\langle m\rangle^2 \geq 0
\end{equation}
if $\rho$ is non-negative.  
 At appropriately large values of $t$ the consequences of the Schwartz inequality for a non-negative spectral function are that a plot of $C(t)$ should be concave upward, or equivalently that $M_{eff}$ should fall monotonically with time.  

A qualitative violation of any of these properties is enough to demonstrate non-positivity of $\rho$.  
The lattice evidence, reviewed below, shows that   these qualitative features of $C(t),M_{eff}(t)$ are violated, and in consequence the spectral function must be negative in some places.

\subsubsection{Discovering superconvergence on the lattice}

For $t\gg a$ one can extract a good approximation to the continuum effective mass, and this is what is conventionally studied to reveal non-positivity.  But to reveal superconvergence one must study the regime of small $t$, and   complications arise.  

In principle, there is an easy road to superconvergence, requiring only the measurement of $-\dot{C}(0)$:   
\begin{equation}
\label{ceqn}
 -\frac{dC(0)}{dt}=\frac{1}{2}\int\,\mathrm{d}m^2\,\rho (m^2).
\end{equation}
   In practice this is not so easy, because the lattice effective mass $M_{eff-l}$ is not just $a$ times the continuum effective mass; the dependence on the lattice spacing is more complicated.   To test for superconvergence on the lattice requires some knowledge of the behavior of $\rho$ at large $m^2$, because this governs the relation of $M_{eff-l}$ to the continuum effective mass.

Fortunately, because of AF, we know the needed behavior.  Write the spectral function as a sum of an IR part and a UV part:
\begin{equation}
\label{specfuncdef}
\rho (m^2)=\theta (M^2-m^2)\rho_{IR}(m^2)+\theta (m^2-M^2\rho_{UV}(m^2);
\end{equation}
here $M$ is a judiciously-chosen mass   separating the IR and the UV.  There are corresponding definitions of $C_{UV}(t),C_{IR}(t)$.     From (\ref{ceqn}) we have $\dot{C}_{IR}(t=0)=-\dot{C}_{UV}(t=0)$ if there is superconvergence.

 Take $M$ large enough so that the lowest-order renormalization-group results describe the UV.  Normalize so that the propagator behaves like $1/(p^2 [\ln (p^2/\Lambda^2)]^A$ at large momentum, where (from Sec~\ref{afnonp})$A$ is 13/22 for the Landau-gauge propagator and $A$ = 1 for the PT-RGI propagator.  Using Eq.~(\ref{specfunc}),
\begin{equation}
\label{cuv}
\dot{C}_{UV}(t)= 2^{-1-A}\pi A \int_M^{\infty}\,\frac{\mathrm{d}m}{m}\,\frac{e^{-mt}}{[\ln (m/\Lambda)]^{1+A}}
\end{equation}.  

\begin{figure}
\includegraphics[width=3in]{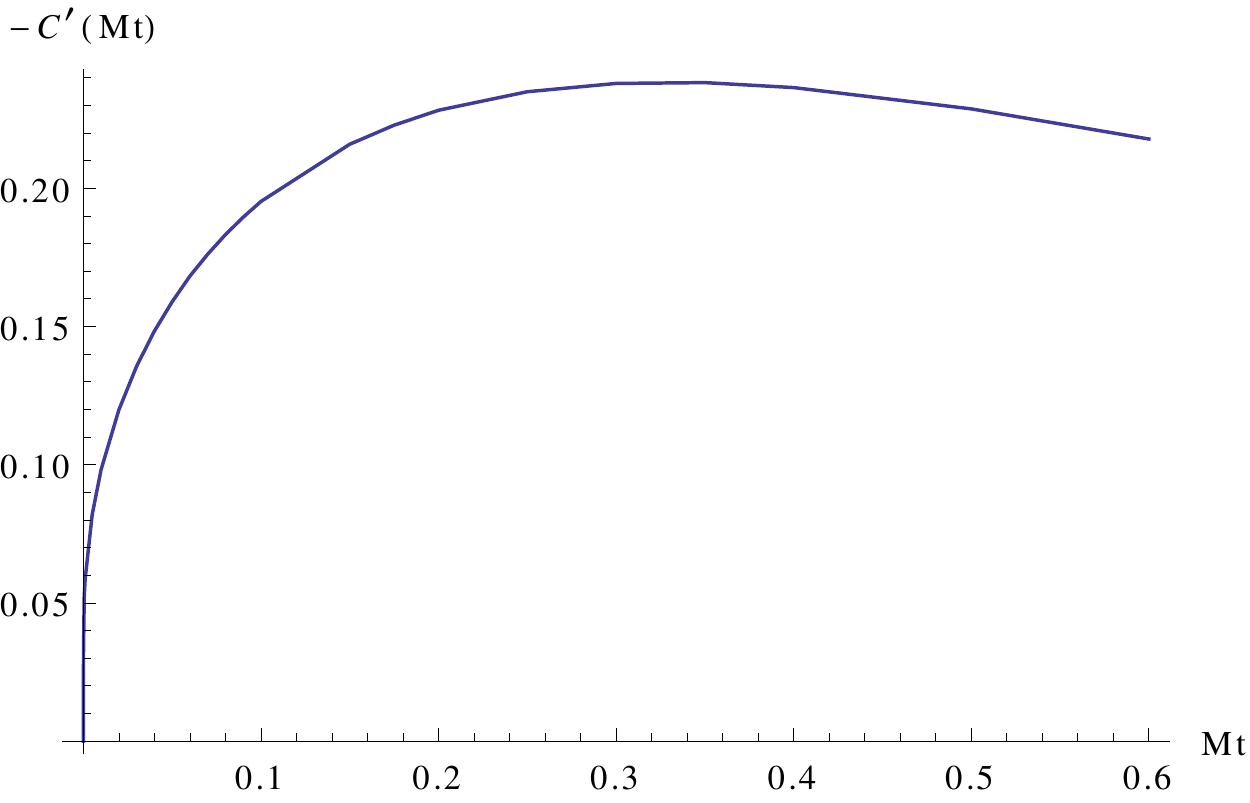}
\caption{A plot of $-\dot{C}(t)$, valid for small $t$, from superconvergence and the   UV spectral function  of Eq.~(\ref{cuv}).  \label{c-fig}}
\end{figure}

This integral converges at $t=0$, thereby setting the value of $\dot{C}_{IR}(t=0)$ if superconvergence is to hold.  However, its derivative is singular at $t=0$.  This makes it somewhat awkward to find the behavior of $\dot{C}(t\approx 0)$, but it turns out that for the full $C(t)$:
\begin{equation}
\label{tnearzero}
-\dot{C}(t) \approx \frac{\pi}{2^{1+A}[\ln (1/(\Lambda t)]^A}[1+\mathcal{O}(1/\ln (1/t)]+\mathcal{O}(t)
\end{equation}
where the $\mathcal{O}(t)$ terms come both from $C_{IR}$ and from non-leading terms of $C_{UV}$.
Fig.~\ref{c-fig} shows the small-$t$ behavior of $-\dot{C}(t)$ for small $t$, based on superconvergence and the asymptotic  behavior in Eq.~(\ref{tnearzero}) for the PT spectral function ($A=1$).

\subsubsection{A special case}

Generally, if there is positivity violation $C(t)$ and its derivatives may have zeroes (other than the superconvergence zero, if any, at $t=0$).  
We now supply a constraint on a specific form of positivity violation proposed by several authors that is useful if superconvergence holds; this form does not give rise to zeroes in $\dot{C}(t)$ other than the superconvergence zero at the origin. 
 Suppose that the spectral function $\rho (\sigma )$ can be divided into two pieces, each of a single sign:
 \begin{equation}
 \label{twopiece}
 \rho (\sigma )=  \rho_{IR} (\sigma )\theta (M^2-\sigma) + \rho_{UV} (\sigma )\theta (\sigma - M^2)
 \end{equation} 
 with $\rho_{IR}\geq 0,\;\rho_{UV}\leq 0)$.  We further suppose that $\rho $ obeys the superconvergence relation
 \begin{equation}
 \label{screl}
 \int_0^{\infty}\, \mathrm{d}\sigma \,\rho (\sigma )=0,
 \end{equation}
 and that $\rho$ is continuous, integrable at infinity, and smooth.  Here $M$ is the location of a zero in $\rho$.  The constraint
 $\rho_{IR}\geq 0 $ is a simple way of ensuring that the gluon propagator obeys $\Delta (p^2=0)> 0$, as all lattice simulations show.  Depending
 on the model of the spectral function, the separating mass $M$ could be a threshhold such that $\rho_{IR}$ has only a simple mass term of  the form
 $\delta (\sigma -M_G^2)$ and is zero otherwise, or models the separation of the $IR$ and $UV$ regions.   Its actual value is immaterial.
 
 Superconvergence implies
 \begin{equation}
 \label{screl2}
 \int_0^{M^2}\, \mathrm{d}\sigma\,\rho_{IR}(\sigma )=K=-\int_{M^2}^{\infty}\, \mathrm{d}\sigma\,\rho_{UV}(\sigma );
 \end{equation}
 the actual value of the positive constant $K$ will not be needed.
 
 Use the language of the wall-to-wall correlator and ask whether   $-\dot{ C}(t)$ can have a zero other than at $t=0$ where it has a zero due to superconvergence.  For a non-negative spectral function it cannot, so zeroes might be indicative of non-positivity.
By definition
 \begin{equation}
 \label{cdot}
 -\dot{C}(t)= \int_0^{\infty}\,m\, \mathrm{d}m\,\rho (m^2)e^{-mt};\;-\dot{C}(t=0)=0.
 \end{equation}
 For any strictly positive $t$ it is elementary that
 \begin{equation}
 \label{ineq}
 -\dot{C}_{IR}(t)\geq Ke^{-Mt};\;|\dot{C}_{UV}(t)|\leq Ke^{-Mt}
 \end{equation}
 where $-\dot{C}(t)_{UV,IR}$ are defined in terms of the corresponding spectral functions.  This shows that the negative contribution of $-\dot{C}_{UV}(t)$ can never cancel the positive IR contribution, for finite $t$; there is only the superconvergence zero.  So for finite $t$, $C(t)$ is monotonic.

\section{\label{lattdata} Lattice data}

Various lattice studies lead to one of the two most widely-claimed properties of the Landau-gauge gluon propagator on the lattice: Its spectral function has a negative part\cite{iritani,gupta,bernard,leinetal,furui,cucch,sternetal,bowman,dudal}. (The other property is that the gluon has a dynamical mass.)  We review some of these data here, and suggest some new methods of analysis of the data that might be useful.  

An important point to start with is that two apparently excellent fits  to a gluon propagator may give rise to very different spectral functions, and that some fits to the propagator cannot even be represented in the spectral form of Eq.~(\ref{klrep}).  In fact, it is a tricky mathematical problem to derive an accurate approximation to spectral function from uncertain knowledge, only at real Euclidean momentum, of a propagator, as of course is the case for lattice data.  The reason\cite{dos,bcdos,dos2} is that to find the spectral function requires inverting a Laplace transform, where it is well-known that a slight error in the function being inverted can lead to a large error in the inversion.  In numerical studies, such errors arise from, among other sources, UV and IR cutoffs and algorithms for doing numerical integrations. In addition, there  are technical features of lattice simulations that can, by themselves, produce apparent non-positivity.  These are lattice artifacts\cite{luscher,aubin} that affect the UV region  (small $t$), but seem not to be important for the large-$t$ IR behavior.  They arise from the use of so-called improved actions, or from lattice implementation of the Landau gauge.  Some authors mention these artifacts, and others do not.  We have no way of knowing, for any set of lattice data, at what values of $t$ the artifacts might or actually do arise.  In discussing various lattice results we therefore must assume that the authors have possible lattice artifacts under control.

At all but the smallest momenta, where Gribov-copy effects can be important, there is wide agreement on the Landau-gauge gluon propagator itself from simulations. However, different groups give quite different interpretations and corresponding proposed spectral functions.  We give three examples.  (There are many more simulations of the Landau-gauge propagator itself; see, {\em e.g.}, Bogolubsky {\em et al.}\cite{bogil}; we consider only examples with no quarks.)  

\begin{figure}
\includegraphics[width=3in]{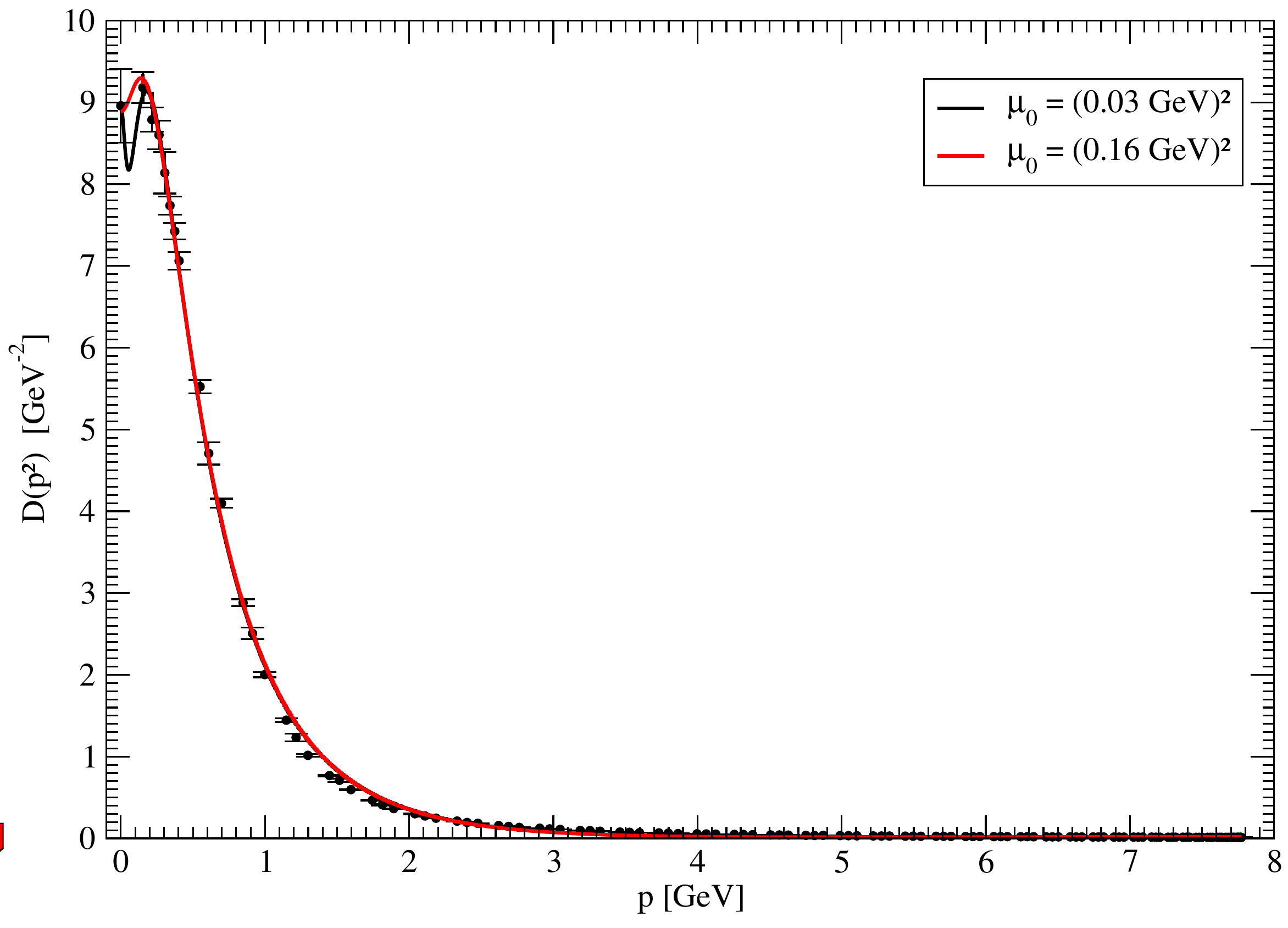}
\caption{ Lattice data points (from \protect\cite{olsil}) and propagators reconstructed from inferred spectral functions (from \protect\cite{dos2}).  \label{olsil-fig}}
\end{figure}

\begin{figure}
\includegraphics[width=3in]{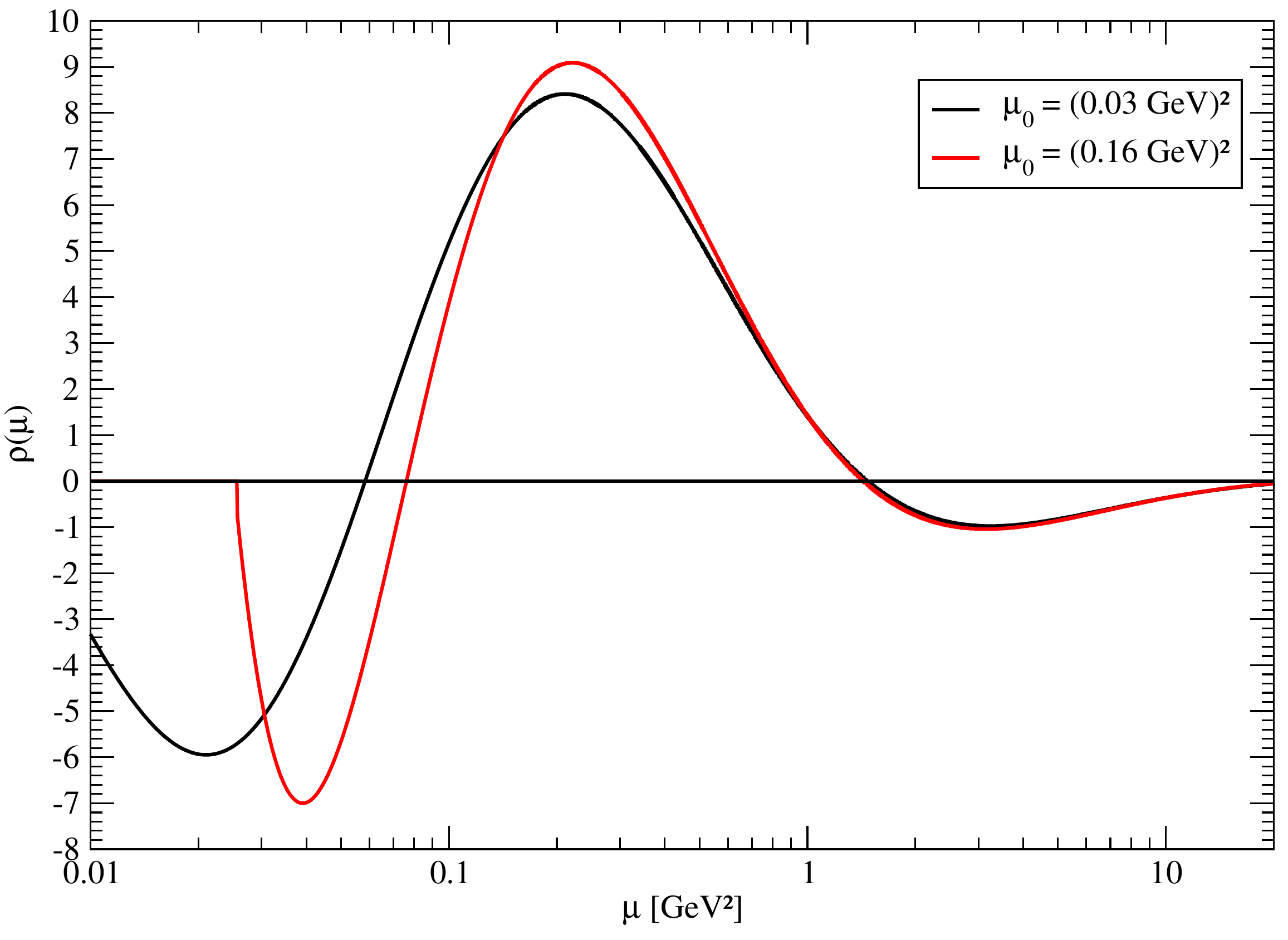}
\caption{  Spectral functions  calculated in Ref.~cite{dos2}.)\label{olsil-fig-2}}
\end{figure}

First, Fig.~\ref{olsil-fig} shows  lattice data points for the propagator\cite{olsil} (along with two fits that we will discuss momentarily).  
The procedure for getting the propagator fits is to estimate the spectral function using methods of Tikhonov and Morozov\cite{dos2}, and then, as a check, to reconstruct the propagator from the spectral function.  Generally, the Tikhonov-Morozov methodology requires introduction of regularization parameters, originally arbitrary and  determined by a minimization principle {\em a posteriori}.  In this case, there is one regularization parameter (a threshhold $\mu_0$ below which the spectral function vanishes) and two minima, each yielding an acceptable spectral function, in principle.  The more regularization parameters the more spectral functions are generated, although  all yield reconstructed propagators \cite{bcdos} agreeing well with data even though the spectral functions differ rather more.
   In the present case the two fits agree well with each other and with the lattice data except at momenta $p\leq 0.2$ GeV, as shown in  Fig.~\ref{olsil-fig-2}.  

\begin{figure}
\includegraphics[width=3in]{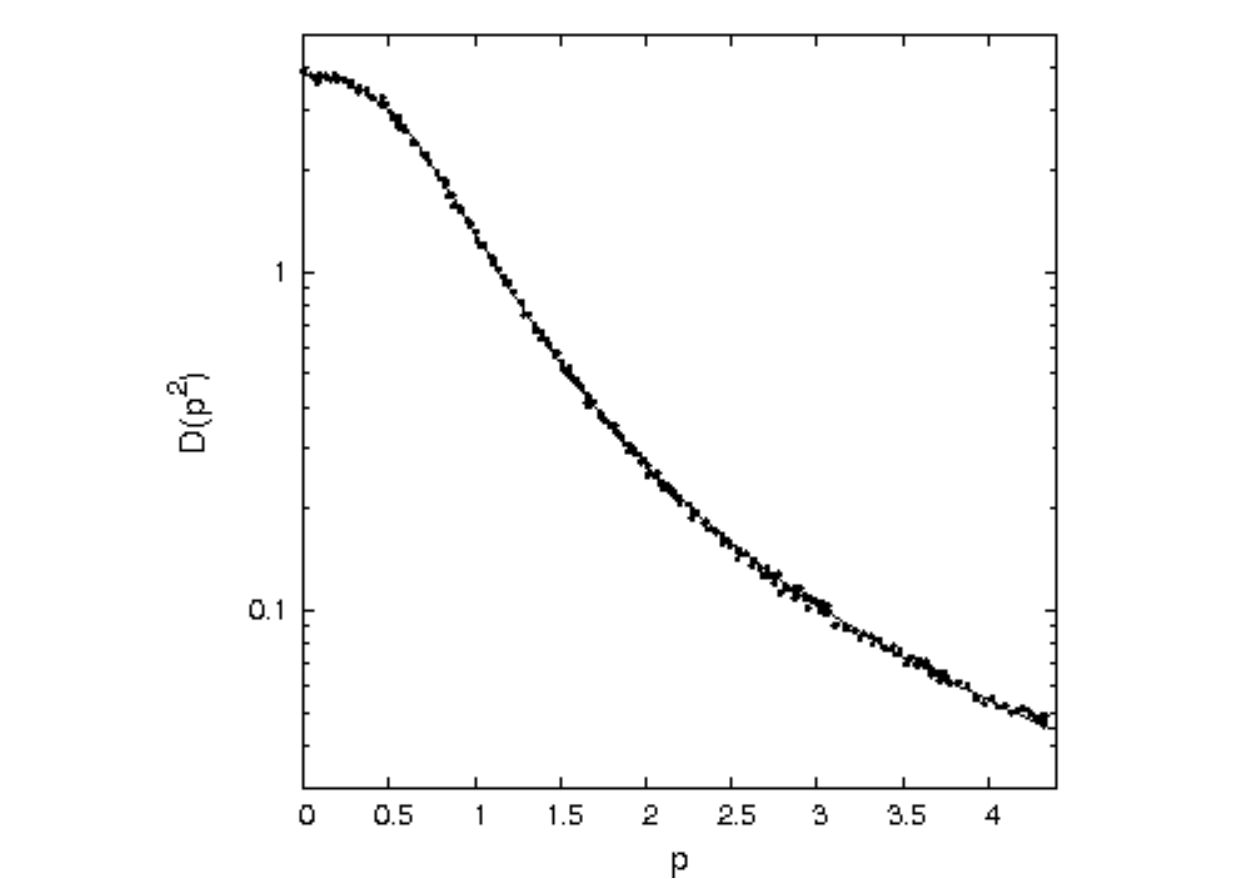}
\caption{ A lattice simulation of the $SU(2)\;d=4$ Landau-gauge gluon propagator, along with a fit function given in the text. (From Cucchieri {\em at al.}\cite{cdmv}.) \label{4d-gluon}}
\end{figure}

The following two examples are not concerned with extraction of the spectral function from the propagator, but instead {\em assume} a spectral function that yields a good fit with the propagator data---in both cases, with negative contributions.   Fig.~\ref{4d-gluon} shows a recent lattice simulation\cite{cdmv} of the $SU(2)\;d=4$ Landau-gauge gluon propagator as a function of momentum $p$, in GeV.  
 The curve is a fit to a propagator of the so-called Gribov-Stingl type, having the form
\begin{equation}
\label{grist}
\Delta (p^2)=const. \frac{p^2+s}{p^4+u^2p^2+t^2}
\end{equation}
with fit parameters $s,t^2,u^2$.  It seems to be an impressive fit, but this propagator form has several defects.  First, it has complex poles, which mean that it cannot be written in the spectral form of Eq.~(\ref{klrep}); second, it does not show AF; and third, it vanishes like $1/p^2$ in the UV, which excludes superconvergence.  Nonetheless, the Cucchieri-Mendes group has analyzed\cite{cucch} propagator data in terms of a spectral function, which has both negative parts and oscillatory parts (coming from the pair of complex-conjugate poles).  

\begin{figure}
\includegraphics[width=3in]{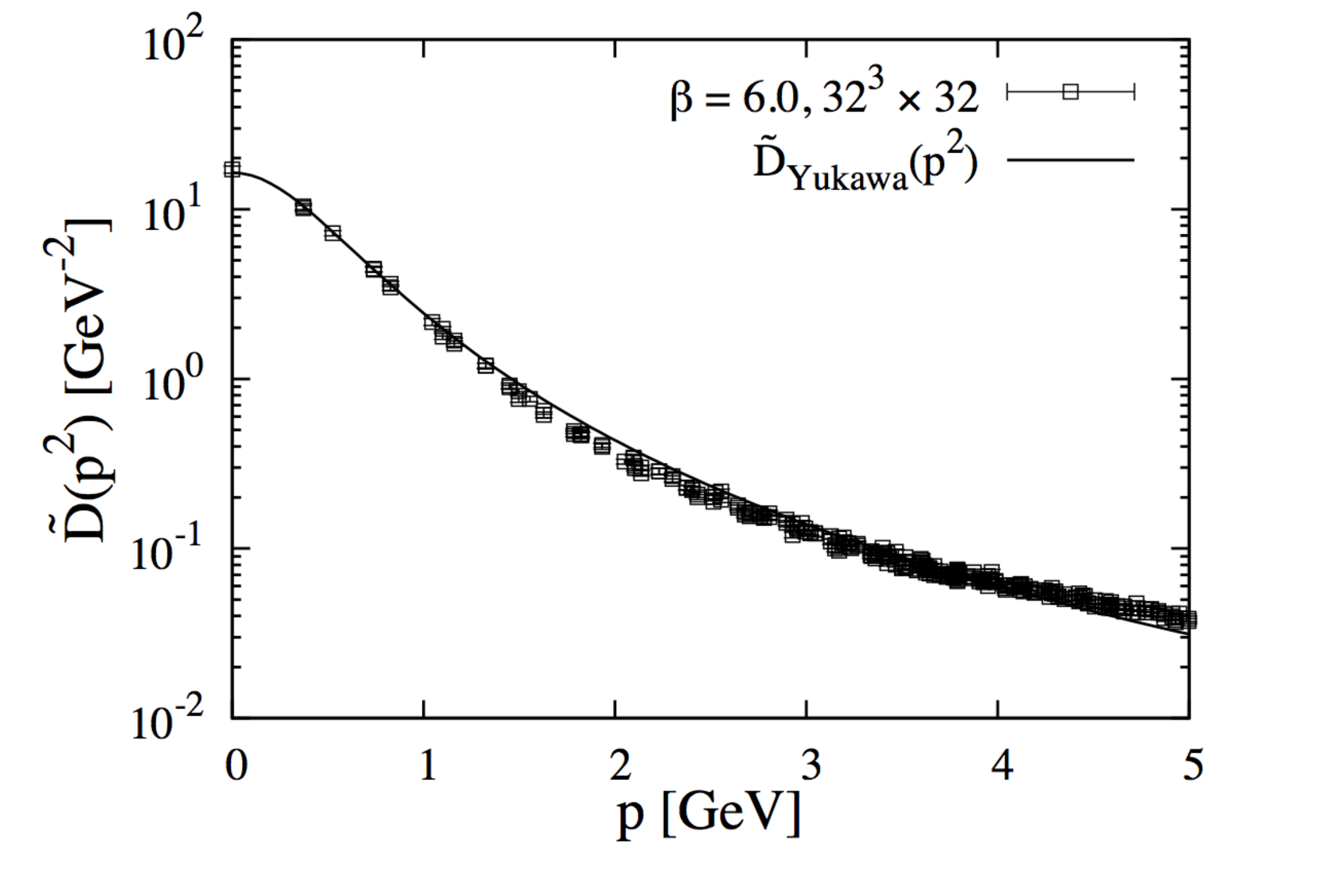}
\vspace*{8pt}
\caption{ A lattice simulation of the $SU(3)\;d=4$ Landau-gauge gluon propagator, along with a fit function given in the text. (From Suganuma {\em et al.}\cite{iritani}.)\label{suganuma}}
\end{figure}

For the  third case, Fig.~\ref{suganuma} shows the $SU(3)\;d=4$ Landau-gauge propagator   from the lattice simulation of a different group\cite{iritani}.
At a glance, the fit, in this case to what the authors call a Yukawa propagator of the form
 \begin{equation}
\label{yukprop}
\Delta (p^2)= const. \frac{m}{(p^2+m^2)^{3/2}}
\end{equation}
is as impressive as the other two examples, but its physical interpretation has to be very different.
This is finite at $p=0$, but of course does not have a conventional mass pole; moreover, it vanishes faster than $1/p^2$ at infinity, and therefore its spectral function satisfies a superconvergence relation.  The authors  claim that a ``regularized" spectral function for this propagator shows a mass-like positive spike  and a continuum spectral function that is everywhere negative.  

The Landau-gauge propagators of Figs.~\ref{olsil-fig}, \ref{4d-gluon}, and \ref{suganuma}  are in  good agreement with each other (up to an overall renormalization constant) except at the smallest momenta, where Gribov-copy and other effects can occur.   There is not even much difference between $SU(2)$ and $SU(3)$  propagators\cite{sternetal,svon,cmos}, in accord with theoretical expectations.    The proposed spectral functions, however,  diverge considerably from each other, both qualitatively and quantitatively.  The   reason   is the lack of a definitive physical interpretation of the propagator dynamics.  

We know of only one first-principles proposal for a superconvergent spectral function that    shows dynamical mass formation; it is based on PT-RGI techniques.

\section{\label{ptsec} The naturalness of non-positivity for the PT-RGI propagator in an AF gauge theory}

We have seen that in an AF NAGT non-positivity of the gluon spectral function is both expected theoretically and found in lattice simulations (in the Landau gauge), yet so far we have presented no interpretation of these results.  It could just as well have happened---perhaps for deep non-perturbative reasons---that lattice simulations showed a non-negative spectral function, for all the understanding we have so far gleaned.  Or possibly these results are somehow peculiar to the Landau gauge.  

However, from the viewpoint of PT-RGI considerations, not only is non-positivity of the spectral function natural and inevitable, it  demonstrates a direct connection between non-positivity and AF\cite{corn076,corn138}.  

\subsection{The renormalization-group-invariant pinch technique}

The renormalization-group-invariant pinch technique (RGI-PT) extracts from an S-matrix element
  off-shell proper $N$-point functions  that are both gauge-invariant and also RGI (that is, independent of a renormalization scale $\mu$)\cite{corn076,cornbinpap,corn138}.   
A longitudinal gluon momentum   arising from gauge-fixing terms or from three-gluon vertices  generate  elementary Ward identities yielding inverse propagators that either vanish (if the line is on-shell) or cancel out propagators in the graph.  The remaining term loses one vertex in the process, resulting  in a graph that is effectively a Feynman graph with one fewer external legs.  When these special graphs are added to normal Feynman graphs with one fewer leg, the result is gauge-invariant (because the S-matrix is). As an example, the result for the one-loop PT (but not RGI) propagator in an $R_{\xi}$ gauge is (we omit an irrelevant $\xi$-dependent gauge-fixing term):
\begin{equation}
\label{fullprop}
\widetilde{\Delta}_{\mu\nu}(p)= P_{\mu\nu}(p)\frac{1}{p^2[1+bg^2\ln (\frac{p^2}{\mu^2})]}+
\dots
\end{equation}
The proper self-energy is gauge-invariant but it depends on the renormalization point $\mu$.
To make this, and other off-shell PT Green's functions, RGI as well as gauge-invariant just multiply the unrenormalized (or renormalized) propagator by the  bare  coupling  $g_0^2$ (or renormalized coupling $g^2$) and divide the appropriate three- and four-gluon vertices by the appropriate coupling\cite{corn138}.  The result for the PT-RGI propagator, independent of $\mu$ and thus RGI, has already been quoted in Eq.~(\ref{ptasym}); note that it is independent of $g^2$.  This works, just as in QED, because of the equality, following from Ward identities, of certain renormalization constants. Based on this one-loop result as phrased in Eqs.~(\ref{runch}, \ref{ptasym2}), we  suggest\cite{corn138} factoring $\widehat{\Delta}$ in the form 
\begin{equation}
\label{corn138}
g^2\tilde{\Delta}(p^2)=\widehat{\Delta}(p^2)=\bar{g}^2(p^2)H(p^2)
\end{equation} 
as the product of the running charge $\bar{g}^2(p^2)$ and a factor $H(p^2)$ that has a form something like an ordinary massive propagator.   As far as has been checked, each factor of this product has a perfectly normal K\"allen-Lehmann representation with a positive spectral function, but their product has a spectral function with negative parts.

\subsection{A phenomenological non-perturbative PT-RGI propagator}

One can illustrate the PT-RGI techniques\cite{corn138} with a simple model extracted from $\phi^3$ theory in six dimensions ($\phi^3_6$, which is well-known to be AF).   
This model is a variant of $\phi^3_6$ in which all particles have the same mass $m$, and two of them carry an 
Abelian charge.    The fundamental idea is to make a good estimate of a gluonic vertex and extract the propagator from a Ward identity.  (Several papers\cite{corn141,corn099,binib} give related investigations of the three-gluon vertex in QCD.) Let $G_{\alpha}(p_1,p_2,p_3)$ be the proper vertex for the Abelian current with momentum $p_{1\alpha}$, divided by $g^2$ (the $\phi^3$ coupling), and $\widehat{\Delta} (p)$ be $g^2$ times the propagator, taken to be the same for all three $\phi$ particles.  Impose the Ward identity
\begin{equation}
\label{phiwi}
p_{1\alpha}G_{\alpha}(p_1,p_2,p_3)=\widehat{\Delta}^{-1}(p_2)-\widehat{\Delta}^{-1}(p_3),
\end{equation}
constructed in analogy to the PT-RGI Ward identity. There are two kinematic structures in $G_{\alpha}$, of which   only one is important for us:
\begin{equation}
\label{vertstruct}
G_{\alpha}(p_i)=(p_3-p_2)_{\alpha}G(p_i)+\dots
\end{equation}
For the propagator in this equation we make the approximation
\begin{equation}
\label{simplerule}
G(p_i)\Delta (p) \rightarrow \frac{1}{p^2+m^2}
\end{equation}
and replace dressed vertices in the loop by their bare counterparts.  Provided that the mass $m$ is non-running this {\em ansatz}  automatically satisfies Ward identities needed for the PT and gives the right UV behavior\cite{corn138}.
  Use these forms in the integral on the right-hand side of the SDE to find, after momentum integration, the Feynman-parameter form:
\begin{equation}
\label{photvert}
G_{\alpha}(p_i)= -3b\int\![\mathrm{d}z]\, \ln [\frac{\Lambda^2}{D+m^2}]
 [p_2(1-2z_3)-p_3(1-2z_2)]_{\alpha}.
\end{equation}
where $\Lambda$ is the {\em physical} mass of the theory and
\begin{eqnarray}
\label{zint}
\int\![\mathrm{d}z] & = & 2\int_0\!\mathrm{d}z_1\,\int_0\!\mathrm{d}z_2\,\int_0\!\mathrm{d}z_3
 \,\delta (1-\sum z_i),\\
D & = & p_1^2\,z_2z_3+p_2^2\,z_3z_1+p_3^2\,z_1z_2 .
\end{eqnarray}
Note that the   Born term has been completely subsumed in  replacing the UV cutoff of the loop integral $\Lambda_{UV}^2$ by the physical mass $\Lambda^2$.  The PT-RGI propagator comes from the Ward identity:
 \begin{equation}
\label{solvewi}
\Delta^{-1}(p_3)=6b\int\!\mathrm{d}z_1\,\mathrm{d}z_2\,\delta (1-z_1-z_2)[D(z_3=0)+m^2]\ln [\frac{D(z_3=0)+m^2}{e\Lambda^2}].
\end{equation} 
Because $D\sim p^2$ at large momentum, we can see that AF demands non-positivity:  This propagator decreases in the UV as $1/(p^2\ln p^2)$ and therefore does not have a positive spectral function.  But one can factor it as in Eq.~(\ref{corn138}), with one of the factors being the running charge.

\subsection{An approximation to the PT-RGI gluon propagator}

In this section only, we use the Minkowski metric $q^2=q_0^2-\vec{q}^2$ so that timelike vectors have positive norm.

    The basic idea \cite{corn138} is to keep only the one-dressed-gluon loop in the gluon proper self-energy SDE, incorporating  the effects of omitted graphs, such as seagulls and ghost loops, with subtraction constants ($\tilde{Z}, C$ below) that are chosen to meet self-consistency requirements.  Just as for the $\phi^3_6$ model, the one-dressed-gluon loop depends only on the product $G\widehat{\Delta}$, which is replaced by (indices omitted)
\begin{equation}
\label{minkapprox}
G\widehat{\Delta}\rightarrow \frac{G_B}{q^2-m^2+i\epsilon}
\end{equation}
where $G_B$ is the Born vertex, and the mass $m^2$ is non-running.  (From now on we omit writing the $i\epsilon$.)  This is equivalent to the approximation 
\begin{equation}
\label{ptrgiprop2}
\widehat{\Delta}(q^2)=\frac{\bar{g}^2(q^2)}{q^2-m^2} 
\end{equation}
   As above, the UV cutoff appearing in the bare coupling cancels with a cutoff term in the loop integration.  After considerable manipulation of indices, the result is:
\begin{equation}
\label{finiteprop}
\widehat{\Delta}(q^2)^{-1}=q^2b\tilde{Z}+b(q^2+\frac{m^2}{11})J(q^2;\xi )+C
\end{equation}
where
\begin{equation}
\label{jeqn}
J(q^2;\xi )= \int_0^1d\alpha \ln \{\frac{m^2-\alpha (1-\alpha )q^2-i\epsilon
}
{\xi}\}=-q^2\int_{4m^2}^{\infty}\frac{d\sigma}{\sigma}\sqrt{1-\frac{4m^2}
{\sigma}}\frac{1}{\sigma -q^2 -i\epsilon}+\ln (\frac{m^2}{\xi}).
\end{equation}
With the choice $\xi = e^{-2}\Lambda^2$ one can set $\tilde{Z}=0$ so that the UV-asymptotic PT-RGI propagator has the usual form
\begin{equation}
\label{lambda}
\widehat{\Delta}^{-1}(q^2)\rightarrow bg^2q^2\ln (\frac{-q^2}{\Lambda^2})[1+o(1)].
\end{equation} 
The propagator is to have a pole at $q^2=m^2$, which specifies $C$.  Then the final finite form of the PT-RGI inverse propagator is
\begin{equation}
\label{cornhou3}
\widehat{\Delta}(q^2)^{-1}= b\{J(q^2;\xi )(q^2+\frac{m^2}{11})
-J(m^2; \xi )\frac{12m^2}{11}\}.
\end{equation}
It is elementary but tedious to recover the spectral density $\rho$ from this expression; it has\cite{corn138} several features that fit into previous discussion.  First, it obeys superconvergence.  Second, it is the sum of one positive and one negative spectral function, with the negative function, at large $\sigma$, of the form in Eq.~(\ref{specfunc}) (with $A=1$).  

It remains to be seen at what level of quantitative accuracy the proposal of this section, or any other proposal, fits the lattice data, since the needed data analysis is yet to be done.  This analysis is not easy since it will require comparing data at a sequence of smaller and smaller lattice spacings (or smaller and smaller $t$ in the wall-to-wall correlator $C(t)$) to look for superconvergence and for the singular derivatives of $C(t)$ expected from AF.

\section{Outlook}

Lack of positivity for the spectral function of the gluon propagator in the Landau gauge is well-established qualitatively, but simulations so far reveal little if anything about the underlying physics, even qualitatively, because of the loss of accuracy in inverting a Laplace transformation with errors. This suggests continued direct study, as outlined in Sec.~\ref{analysis}, of the function $C(t)$   in order (for example) to establish superconvergence.   Moreover, the Landau-gauge propagator is not a physical object, which suggests more effort to reconstruct the PT propagator,  perhaps along the lines of Sec.~\ref{ptprop}. These are challenging problems for both theorists and simulators.

\begin{acknowledgments}

I am happy to acknowledge informative correspondence with Paulo Silva about his and his collaborators' work on positivity.

\end{acknowledgments}

\end{document}